\documentclass[12pt,preprint]{aastex}
\def\gtwid{\mathrel{\raise.3ex\hbox{$>$\kern-.75em\lower1ex\hbox{$\sim$}}}}
\def\ltwid{\mathrel{\raise.3ex\hbox{$<$\kern-.75em\lower1ex\hbox{$\sim$}}}}
\def\\{\hfil\break}
\def\ie{{\it i.e.\ }}
\def\eg{{\it e.g.\ }}
\def\etal{{\it et al.\ }}

\begin{document}
\title{Optimal Moments for the Analysis of Peculiar Velocity Surveys II:
Testing}
\vskip 0.5cm
\author{\bf Hume A. Feldman$^{\dagger\star}$\footnote{feldman@ku.edu},\, Richard
Watkins$^\ddagger$\footnote{rwatkins@willamette.edu}, Adrian
L. Melott$^\dagger$\footnote{melott@kusmos.phsx.ukans.edu} \, \, \&
Scott W. Chambers$^\dagger$\footnote{willc@ku.edu}}
\vskip 0.3cm
\affil{{\it $^\dagger$Department of Physics \& Astronomy, University of
Kansas, Lawrence, KS 66045, USA}}
\affil{\it $^\star$Racah Institute of Physics, Hebrew University, Jerusalem, 91904, Israel}
\affil{\it $^\ddagger$Department of Physics, Willamette University,
Salem, OR 97301, USA}

\baselineskip 12pt plus 2pt

\begin{abstract}

Analyses of peculiar velocity surveys face several challenges, including
low signal--to--noise in individual velocity measurements and the
presence of small--scale, nonlinear flows.  This is the second in a
series of papers in which we describe a new method of overcoming these
problems by using data compression as a filter with which to separate
large--scale, linear flows from small--scale noise that can bias
results.  We demonstrate the effectiveness of our method using realistic
catalogs of galaxy velocities drawn from N--body simulations.  Our tests
show that a likelihood analysis of simulated catalogs that uses all of
the information contained in the peculiar velocities results in a bias
in the estimation of the power spectrum shape parameter $\Gamma$ and
amplitude $\beta$, and that our method of analysis effectively removes
this bias.  We expect that this new method will cause peculiar velocity
surveys to re--emerge as a useful tool to determine cosmological
parameters.

\end{abstract}

\noindent{\it Subject headings}: cosmology: distance scales -- cosmology: large
scale structure of the universe -- cosmology: observation -- cosmology:
theory -- galaxies: kinematics and dynamics -- galaxies: statistics

\baselineskip 23pt plus 2pt

\section{INTRODUCTION}
\label{sec-intro}

Although in principle the measurement of galaxy motions holds great
promise as a probe of large--scale structure, in practice there are
several obstacles that have prevented there from being robust
conclusions made from the analyses of these measurements.  First, these
measurements are inherently noisy; errors in peculiar velocity
determinations are typically of order $10\%$ of the redshift of a galaxy
or cluster, which for all but nearby objects is comparable or larger
than the velocity being measured.  From the theoretical side, in order
to relate the velocity field to the underlying matter density one must
assume that the fields are linear.  While this approximation is accurate
on large scales, on smaller scales it generally fails due to infall into
density concentrations.  The difficulty in surmounting these obstacles
is illustrated in the fact that attempts to compare different velocity
field surveys have shown significant disagreements
\citep{wf95,hudson99}.   

Analyses of catalogs of peculiar velocity measurements have usually
taken either of two main approaches. One is to average all the
velocities to find the bulk flow, the velocity of the volume occupied by
the survey relative to the Universal rest frame defined by the CMBR (\eg
\citet{LP,RPK,branchini,colless,AG01}). This method has
the disadvantage that it discards most of the information contained in a
survey and measures only three quantities, the components of the bulk
flow vector.  The second approach is to use all of the information
contained in the survey in a likelihood analysis in order to obtain
maximum likelihood estimates of the power spectrum parameters (\eg
\citet{ENEARPS,yang01,zehavi,SFI}).  This method potentially
suffers from a bias due to small--scale, nonlinear contributions to
velocities.

In a recent paper (\citet{paperI}, hereafter Paper I), we introduced a
new method for the analysis of peculiar velocity surveys that is a
significant improvement over previous methods.  In particular, our
formalism allows us to separate information about large--scale flows
from information about small scales, the latter which can then be
discarded in the analysis.  By applying specific criteria, we are able
to retain the maximum information about large scales needed to place the
strongest constraints, while removing the bias that small scale
information can introduce into the results.

In paper I we reported on preliminary tests that suggested that our
analysis method was effective.  Here we present results from more
extensive testing that demonstrates conclusively that our method works
as advertised.  In particular, we show that a likelihood analysis of
simulated catalogs that uses all of the information contained in the
peculiar velocities results in a bias in the estimation of the power
spectrum parameters $\Gamma$ and $\beta$, and that our method of
analysis effectively removes this bias.

The paper is organized as followed: in sections 2 and 3 we briefly
review the analysis formalism introduced in Paper I. In section 4 we
discuss the N--body simulations we used and how we constructed synthetic
catalogs. In section 5 we discuss results from our synthetic catalogs
for both a full likelihood analysis of the catalog as well as our new
analysis method.  We also present additional evidence that our method is
working effectively.  In section 6 we conclude.

\section{The Formalism}
\label{sec-formalism}

Our starting point is the usual statistical model for the
line--of--sight peculiar velocities of galaxies.  First, we assume that
galaxies are tracers of a large--scale linear velocity field ${\bf v(\bf
x)}$, a Gaussian random field completely described by the velocity power
spectrum $P_{(v)}(k)$.  With the assumption of linearity, the velocity
power spectrum is proportional to the power spectrum of density
fluctuations, with $P_{(v)}(k)= H_o^2\Omega^{1.2} P(k)$.

For a set of $N$ galaxies with positions ${\bf r_i}$, the observed
line--of--sight velocity will be given by
\begin{equation}
v_i = {\bf v}({\bf r}_i)\cdot {\bf \hat r}_i + \delta_i,
\label{vo}
\end{equation}
the sum of the radial component of the velocity field with a noise term
$\delta_i$ describing both observational error and any deviation from
the linear velocity field due to local gravitational interactions.  For
simplicity we assume that $\delta_i$ is distributed as a Gaussian random
variable with variance $\sigma_i^2 + \sigma_*^2$, where $\sigma_i$ is
the observational error associated with that particular galaxy, and
$\sigma_*$ describes all other effects and is assumed to be the same for
all the galaxies in the set.

With these assumptions, we can construct the probability distribution
for the set of measured line--of--sight velocities given a power
spectrum $P(k)$,
\begin{equation}
L(v_1,...,v_N; P(k))= \sqrt{|R^{-1}|}\ \exp\left(\sum_{i,j=1}^N -v_i
R^{-1}_{ij} v_j/2\right),
\label{like}
\end{equation}
where $R_{ij}= \langle v_{i}\ v_{j}\rangle $ is the covariance matrix,
which in this case takes the form
\begin{equation}
        R_{ij} = R^{(v)}_{ij} + \delta_{ij}\, (\sigma_{i}^{2}+
        \sigma_{*}^{2})
\label{Rij}
\end{equation}
where $R^{(v)}_{ij} = \langle {\bf v}({\bf r}_{i})\cdot {\bf
\hat r}_{i} \ \ {\bf v}({\bf r}_{j})\cdot {\bf \hat
r}_{j}\rangle $ and the second diagonal term is due to the noise.  In
linear theory, the ``signal'' part of the covariance matrix $R^{(v)}$
can be written as an integral over the density power spectrum
\begin{equation}
R^{(v)}_{ij} = {H^2\Omega_o^{1.2}\over 2\pi^2}\int P(k)W^2_{ij}(k)\ dk,
\label{Rvij}
\end{equation}
where $W^2_{ij}(k)$ is a tensor window function calculated from the set
of positions ${\bf r}_i$ of the galaxies with galaxy velocities weighted
by their relative error (for more details see \citet{fw94,wf95,fw98}).

Typically we are given the a catalog of measured velocities
$(v_1,...,v_N)$ and wish to determine $P(k)$.  Thus we can view
$L(v_1,...,v_N;P(k))$ as a likelihood functional; given $(v_1,...,v_N)$
we use Eq. (\ref{like}) to determine the likelihood that they were
generated in a Universe with a particular $P(k)$.

An analysis of the type described above is susceptible to biases due to
small--scale, nonlinear contributions to the galaxy velocities.  The
method of analysis developed in Paper I eliminates this bias by
replacing the full set of $N$ line--of--sight velocities $v_i$ with
moments $u_i$, which are designed so that the $N$th moment is the linear
combination that carries the most information about small scales, the
$(N-1)$th moment is an independent linear combination which carries the
second most small--scale information, etc.  By using a subset of the
first $N^{\prime}< N$ moments in our analysis and discarding the rest,
we can essentially filter out the small scale information which may
carry nonlinear contributions.  Our method is based on Karhunen--Lo\`eve
methods of data compression \citep{KK,KS}; see also \citet{TTH},
designed to concentrate most of the information in a large set of data
into a smaller, more manageable number of moments.  However, our method
puts a twist on this idea by concentrating unwanted information
regarding small scales into a small set of moments which can then be
discarded.  What follows is a brief review of the mechanics of our
method; for details, see Paper I.

Our method is based on {\it linear} data compression, so that the moment
$u_n$ can be written in terms of the line--of--sight velocities $v_i$ as
\begin{equation}
u_{n}= \sum_{j=1}^N (b_n)_j\ v_{j}\ ,
\end{equation} 
where the $b_{n}$ are a set of vectors of length $N$.  With this
definition, the covariance matrix of the new moments is given as
\begin{equation}
{\tilde R_{nm}} = \langle u_n u_m\rangle =
\sum_{i,j} (b_n)_i\langle v_iv_j\rangle (b_m)_j = 
\sum_{i,j} (b_n)_iR_{ij}(b_m)_j\ .
\label{Rtilde}
\end{equation}
 It is convenient to choose the $b_n$ so that the $u_n$ are linearly
 independent and of unit variance, so that ${\tilde R_{nm}}$ is the
 identity matrix.  Note that this normalization will hold only for a
 particular matrix $R_{ij}$ and hence a particular power spectrum.

In order to find the vector $b_n$ such that the moment $u_n$ carries the
maximum information about nonlinear scales, we assume a simple model for
the power spectrum in which the amount of power on scales below that
where density fluctuations have gone nonlinear is specified by a single
parameter $\theta_q$.  Given a single moment $u_n$, we can determine the
value of $\theta_q$ to within a minimum variance given by
$\Delta\theta_{q}^{2} = 1/{\tilde F}_{qq}$, where ${\tilde F}_{qq}$ is
the $qq$th element of the Fisher information matrix, which in this case,
and with the normalization assumed above, can be shown to take the form
\begin{equation}
{\tilde F_{qq}}= \sum_{ij} {1\over 2}\left((b_n)_i{\partial
R_{ij}\over\partial\theta_q} (b_n)_j\right)^2
\end{equation}

We can thus find the single moment $u_n$ that carries the maximum
information about $\theta_q$ by finding the vector $b_n$ which minimizes
${\tilde F}_{qq}$ subject to the normalization condition discussed
above, which functions as a constraint.  After introducing a Lagrange
multiplier, the minimization results in an eigenvalue problem
\begin{equation}
\sum_{i,j,m}    \left( L^{-1}_{ki}{\partial R_{ij}\over
\partial \theta_{q}}L^{-1}_{lj}\right) \left(L_{ml}b_{m}\right) =
\sum_j \lambda \left( L_{jk}b_{j}\right)
\end{equation}
where $L_{ij}$ is the Cholesky decomposition of the covariance matrix,
$R_{ij} =\sum_{p=1}^N L_{ip}L_{jp}$.

Solving this eigenvalue problem gives us a set of $N$ orthogonal
eigenvectors $\sum_j L_{ji}(b_n)_j$ with eigenvalues $\lambda_n$.  Each
eigenvector has a corresponding moment $u_n= \sum_i (b_n)_i v_i$.  The
eigenvalue $\lambda_n$ of a moment $u_{n}$ is related to the error bar
$\Delta\theta_{q}$ that one could place on $\theta_{q}$ using the single
moment $u_n$, as can be seen by manipulating the equations above:
\begin{equation}
{1\over
\Delta\theta_{q}^2}= {\tilde F}_{qq} =
\sum_{i,j} {1\over 2}\ (b_i{\partial R_{ij}\over
\partial
\theta_{q}} b_j)^2= \sum_{i,j}{1\over 2}\ 
({ b_i \left(\lambda R_{ij}b_j\right)})^2 = {\lambda^2\over 2}
\end{equation}
so that $\Delta\theta_{q} = 2/ \lambda^2$.  If we order the moments
$u_{n}$ in order of increasing eigenvalue,
\begin{equation}
        |\lambda_{1}|\le |\lambda_{2}|\le \ldots\le |\lambda_{N}|
\label{ordering}
\end{equation}
then we can interpret each moment as carrying successively more
information about $\theta_{q}$, with $u_{N}$ carrying the maximum
possible amount of information.  Since our goal is to produce a data set
that is less sensitive to the value of $\theta_{q}$ than the original
data, we should keep moments only up to some $N^{\prime}$.  The
orthogonality of the eigenvectors ensures that the moments are
statistically independent, as we assumed above.  Thus if we compress the
data by discarding the moments with a large value of $|\lambda|$, the
information contained in those moments will be completely removed from
the data.  However, we would also like to keep as many moments as
possible in order to retain the maximum information about large scales.

In order to choose a value of $N^{\prime}$, we need to examine what
error bar $\Delta\theta_{q}$ we can put on the parameter $\theta_{q}$
using the compressed data.  Since the moments are independent, we can
write the Fisher matrix for the $N^{\prime}$ moments that were not
discarded as
\begin{equation}
\tilde F_{qq}= 
\sum_{n=1}^{N^{\prime}}{1\over 2}\lambda_{n}^{2}
\end{equation}
so that the error bar that can be put on $\theta_{q}$ using the
compressed data is given by
\begin{equation}
        \Delta\theta_{q} = {1\over\sqrt{\tilde F_{qq}}}= \left[ {1\over
        2}\sum_{n=1}^{N^{\prime}}\ \lambda_{n}^{2}\right]^{-1/2}
\label{criteria}
\end{equation}
This suggests that $N^{\prime}$ should be chosen by adding up the sum of
the squares of the smallest eigenvalues until the desired sensitivity is
reached.  The criterion that we use is as follows: First, we estimate
the actual size of the parameter $\theta_{q}=\theta_{qo}$ from peculiar
velocity data.  Then, we keep the largest number $N^{\prime}$ moments
that is still consistent with the requirement that $\Delta\theta_{q}\ge
\theta_{qo}$.  With this requirement, as long as our estimate of the
true value of $\theta_{q}$ is correct, our final set of moments
$u_{1}\ldots u_{N^{\prime}}$ will not contain enough information to
distinguish the value of $\theta_{q}$ from zero.

\section{Other Selection Criteria}
\label{selection}

Our method of selecting moments by their lack of information about small
scales has the disadvantage of not discarding moments which have little
information about any scale; this issue was briefly touched on in Paper
I.  Thus we have developed a second criterion for moment selection
whereby we discard moments that are dominated by noise.  That is, they
have no cosmologically useful information.

Recall from Eq.~\ref{Rij} that the covariance matrix for the
line--of--sight velocities is the sum of a ``signal'' part and a noise
part.  Since the covariance matrix for the moments $u_n$ is essentially
a ``rotation'' of the velocity covariance matrix, this matrix can
separated in a similar fashion,
\begin{eqnarray}
\tilde R_{nm}&=&\sum_{ij} (b_n)_iR_{ij}(b_m)_j=
 \sum_{ij} (b_n)_iR^{(v)}_{ij}(b_m)_j +
\sum_{ij} (b_n)_i\delta_{ij}(\sigma_i^2+\sigma_*^2)(b_m)_j\nonumber\\
&=&
\tilde R_{nm}^{(v)} + \sum_i (b_n)_i(b_m)_i(\sigma_i^2+\sigma_*^2)
\end{eqnarray}
where the second term is the noise contribution to the variance of the
moment.  Given that the $b_n$ are normalized such that the moments are
independent and have unit variance, {\it i.e.}  that $\tilde R$ is the
identity matrix, we see that the quantity
\begin{equation}
\xi_n= \sum_i (b_n)_i^2(\sigma_i^2+\sigma_*^2)
\label{noise}
\end{equation}
is a measure of the fraction of the variance of the moment $u_n$ that is
due to noise.  If $\xi_n\ll 1$, then the moment has very little noise
and should be retained.  If, on the other hand, $\xi_n \approx 1$, then
the value of the moment is mostly determined by the errors in the data
and should be discarded.

Generally there is a correlation between a moment's $\xi_n$ and its
eigenvalue $|\lambda_n|$; moments that are most sensitive to small
scales tend to be very noisy.  Low noise moments tend to be those that
probe large--scale power.  This is due to the fact that the measurement
errors in the velocities, which vary independently from galaxy to
galaxy, are much more effective at masking small scale modes than large
scale modes.  However, we have found that some moments with small
eigenvalues also have large noise; these moments carry little
information about {\it any} scale.

The correlation between $\xi_n$ and $|\lambda_n|$ suggests that our two
selection criterion can be somewhat redundant; eliminating noisy moments
often can also accomplish the goal of removing moments with large
$|\lambda_n|$.  Similarly, eliminating moments with large $|\lambda_n|$
leaves one with moments which generally have smaller noise.  Thus we
will see below that applying the second criteria after we have already
applied the first typically does not change the results of the analysis
significantly.

Once one has a set of moments which have both small $\xi_n$ and
$|\lambda_n|$, it is desirable to have a way of determining which scales
each moment is most sensitive to.  From Eq.~\ref{Rvij} we recall that
the ``signal'' part of the covariance matrix for the velocities
$R^{(v)}$ is given by an integral of a tensor window function
$W_{ij}(k)$ with the power spectrum.  The ``signal'' part of the
covariance matrix $\tilde R$ of the moments is given by a ``rotation''
of this tensor window function.  Since $\tilde R$ is diagonal, this
results in a scalar window function for each moment $u_n$,
\begin{equation}
W^{2}_{n}(k)= (b_{n})_{i} W^2_{ij}(k)(b_{n})_{j}
\label{Wsq} 
\end{equation}
By examining the window functions for each moment, we can determine
which scales the moments are sensitive to and confirm that our method is
working.  In principle, examination of the window functions could also
provide a further criterion for the discarding of moments.

\section{Synthetic Catalogs}
\label{catalogs}

The simulations used here are numerical models for the gravitational
dynamics of collisionless particles in an expanding background. We are
studying evolution of initial Gaussian perturbations in a
matter--dominated universe. All the simulations are done with a
particle--mesh (PM) code with $256^3$ particles in an equal number of
grid points
\citep{m86,MWG88}.  More details about the peculiarities of the
simulations used here can be seen in \citet{ms93}. Although the
parameter $\Omega$ appears in both the dynamics of the expanding
background and as a parameter in the fit to the power spectrum shape in
the CDM family of models \citep{BBKS}, they serve two different
functions.

We ran simulations with a dynamical background $\Omega_0$ = 1. or 0.34.
These were normalized to an amplitude $\sigma_8$=0.93 at redshift moment
$z$=0, but we also took data at $z$=1, which for our purposes can be
described as studying a Universe with a lower perturbation amplitude
normalization, and possibly a higher $\Omega_0$.  All models were
interpreted with an assumed Hubble Constant H$=100h$km
s$^{-1}$Mpc$^{-1}$ where $h$ = 2/3.  When power spectra are
parameterized in $Mpc$, the shape is dependent upon $\Omega_0 h^2$,
which we set equal to 0.15, 0.35, and 1.0 for our low--$\Omega$ tests.
We have used some values which are inconsistent with other constraints
in CDM linear theory in order to test our method over a wide range of
values.  We also ran another simulation with $\Omega_0$ = 1, and
$\Omega_0 h^2$ = 0.15; such models have been called $\tau$CDM in the
past.  The set with $\Omega_0$ = 0.34, $\Omega_0 h^2$ = 0.15 is most
consistent with a variety of findings at this time, but we do not wish
to test our method {\it only} against currently favored cosmologies.
There are a variety of alternative models in addition to the
cosmological constant $\lambda$; all have very small and totally linear
effects on large--scale velocities.  We omit this in favor of wider
exploration of parameter shifts which have large effects.

For testing our method we created synthetic redshift--distance catalogs
from the 256$^{3}$ $N$--body PM simulations.  In these simulations, the
box size was taken to equal 512 Mpc, or 34,133 km s$^{-1}$ in redshift
space for $h$=2/3.

Each of the points defined by the mesh represented a galaxy with a
corresponding location and velocity.  Testing the optimal moments method
accurately requires proper modeling of statistical errors so we wanted
to be sure to include the effects of cosmic variance and scatter in
distance indicators.  We chose three coordinates of scale 1/6, 1/2 and
5/6 the box width.  An exhaustive permutation of combining these
coordinates results in 27 center locations within the box, each
corresponding to an individual synthetic catalog as described next.
About each central location, an annular volume was defined by the
redshift range 500 km s$^{-1} <$ z $<$ 10,000 km s$^{-1}$.  If the
volume intercepted the boundaries of the box then we appealed to the
periodic boundary conditions of the simulation and included galaxies
from the opposite box side.  Within each of the defined regions, we
selected $\approx$ 1000 galaxies under the assumption of a radial
selection function and the additional requirement that there is a zone
of avoidance below the galactic latitude $|b|< 10^{\circ}$.  The
selection function was chosen generically (we ignored galactic
properties) to mimic existing popular redshift--distance surveys (\ie
the SFI survey; \citet{dacosta96}).  The effect of scatter in distance
indicators was replicated by adding a random error to each peculiar
velocity drawn from a Gaussian distribution of width 10\% of the galaxy
redshift distance.

In the end, for each simulation box we have 27 surveys sampling the
simulation that give information about the positions and radial
velocities of the galaxy distribution in some volume. We analyzed these
surveys by using the actual positions (\ie no errors) and by perturbing
the velocities with a 10\% Gaussian error. The value of sampling the
simulation in this way is that we are able to model the effects of
cosmic variance.  The errorbars of the unperturbed catalogs are
predominantly from the cosmic variance, whereas those of the perturbed,
10\% catalogs include the effects of both cosmic variance and the
inaccuracies of the distance indicators.

\section{Results}
\label{results}

\begin{figure}[ht]
\plotone{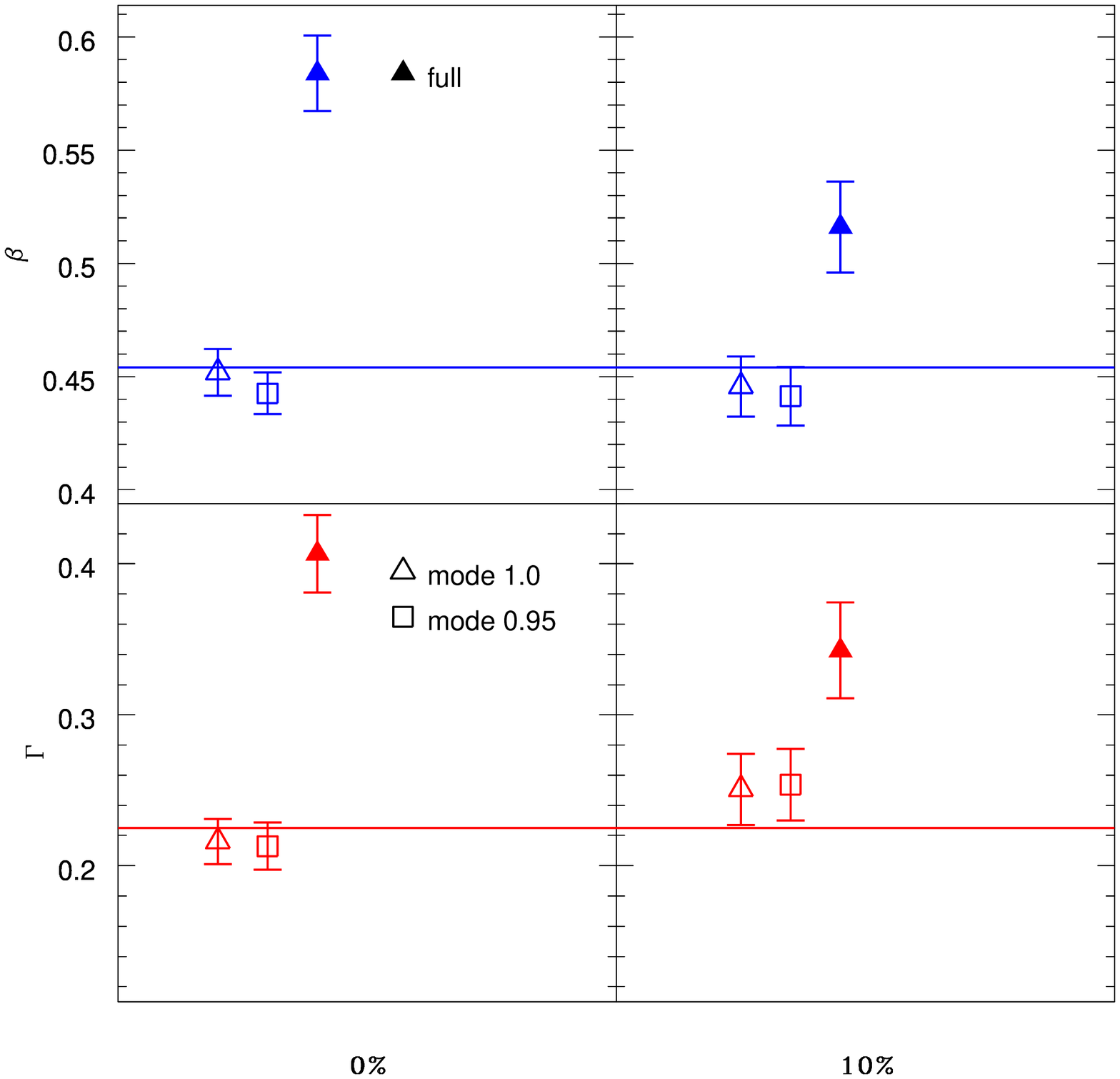}
\caption{A comparison between the mode analysis presented in this
paper and the traditional full analysis. The results are from 81
catalogs extracted from the simulations (see Sec.~\ref{catalogs}). In
the top two panels are the mean values and standard deviations of the
mean of $\beta$, the amplitude of the power spectrum. The bottom panels
we show the results for estimating $\Gamma$, the shape parameter. In the
left panels we have the results for the analysis for a survey with no
errors whereas the right panels show the results for 10\% errors. The
solid symbols are the full analysis results and the empty ones are the
mode analysis. The triangles are the results without removing the noisy
moments, the rectangles are those where we removed the noisiest
moments. The horizontal lines are the ``true'' values of the parameters
($\Gamma=0.225$, $\beta=0.455$). It is clear that the full analysis
fails to recover the parameter values whereas the mode analysis recovers
them well.}
\label{fig-2loz0}
\end{figure}

\begin{figure}[ht]
\plotone{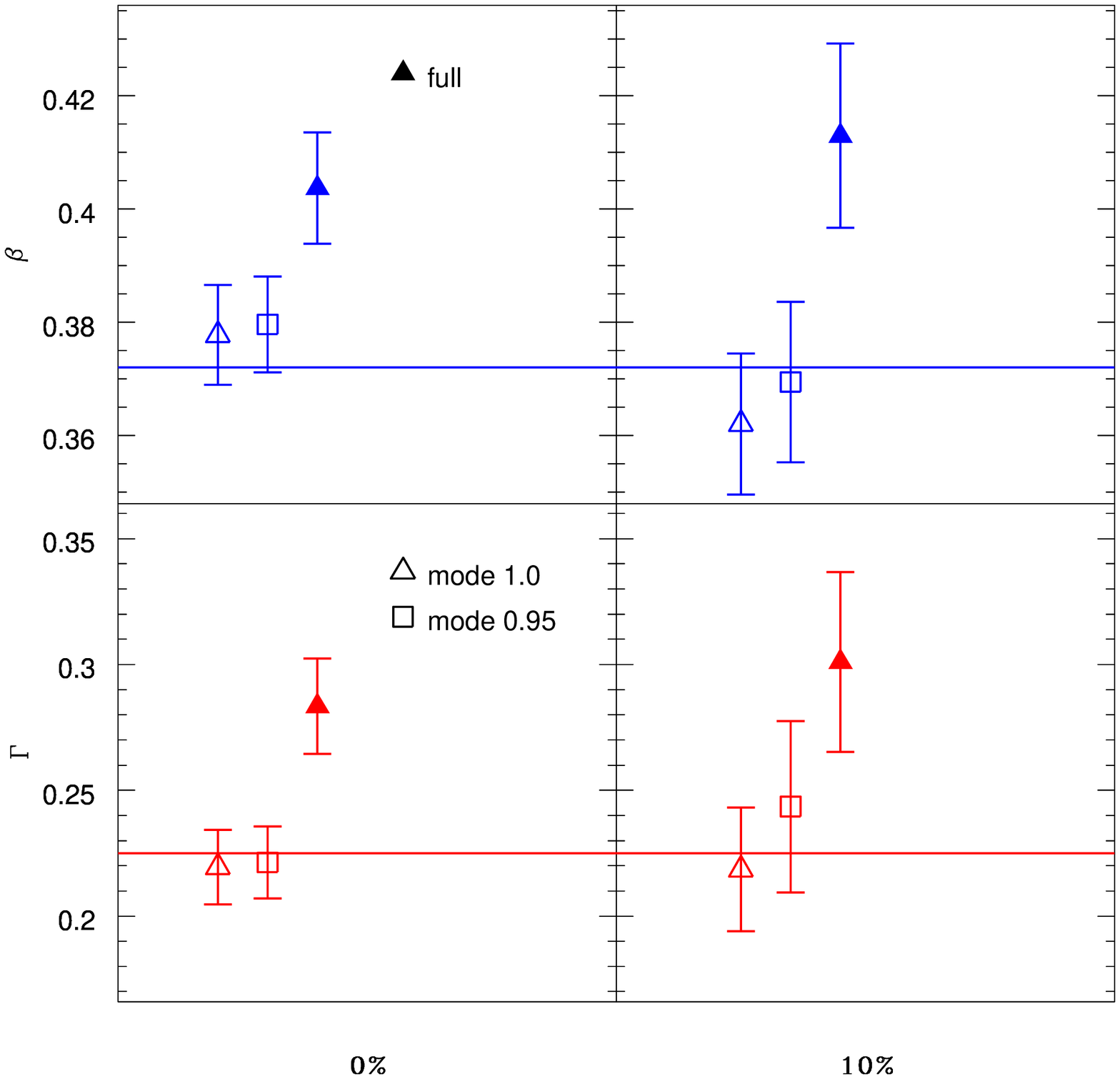}
\caption{
The same as fig~\ref{fig-2loz0} with different cosmological parameters
($\Gamma=0.255$, $\beta=0.372$).}
\label{fig-2loz1}
\end{figure}

\begin{figure}[ht]
\plotone{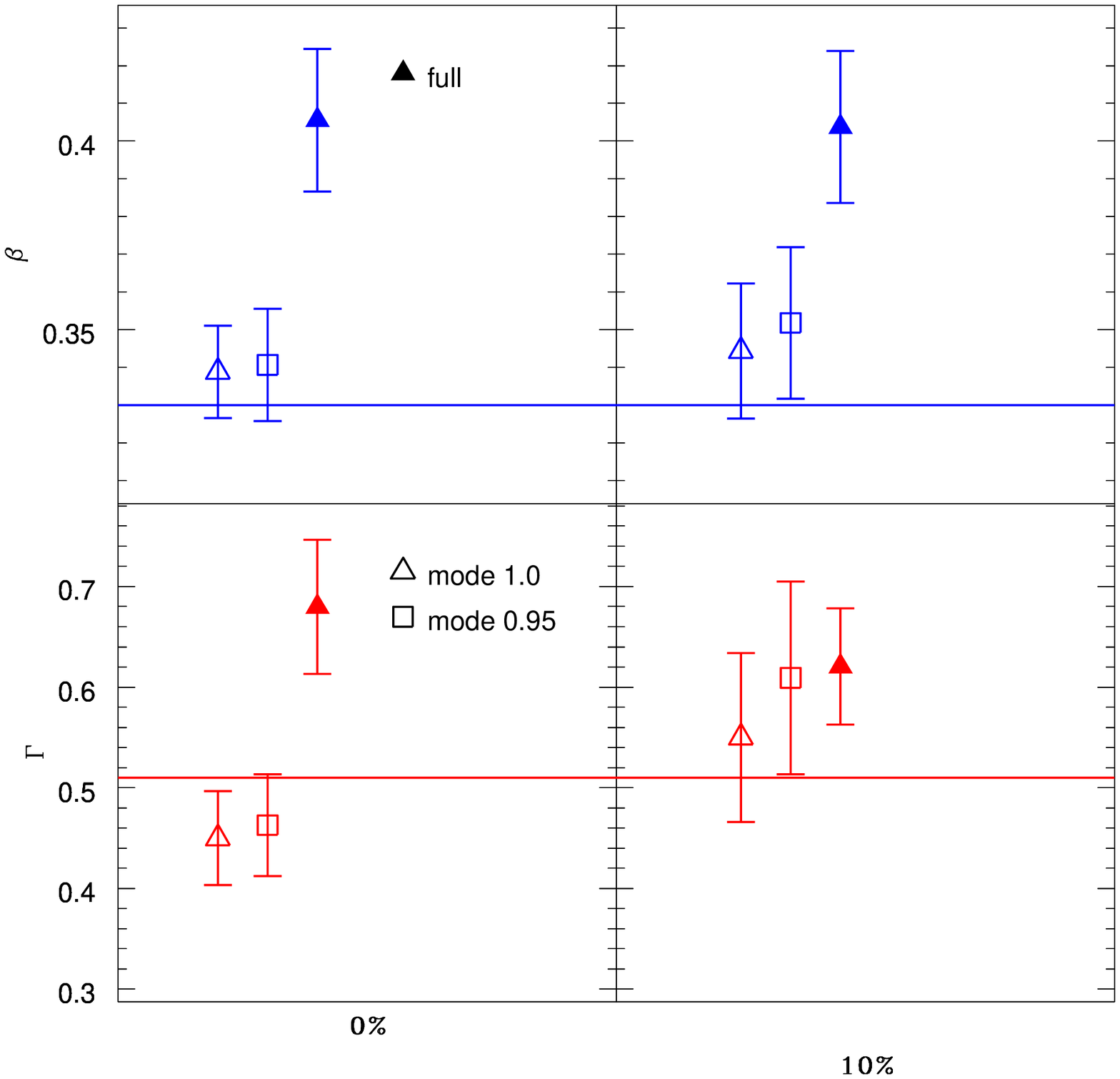}
\caption{
The same as fig~\ref{fig-2loz0} with different cosmological parameters
($\Gamma=0.515$, $\beta=0.325$).}
\label{fig-2lo-35-z1}
\end{figure}

\begin{figure}[ht]
\plotone{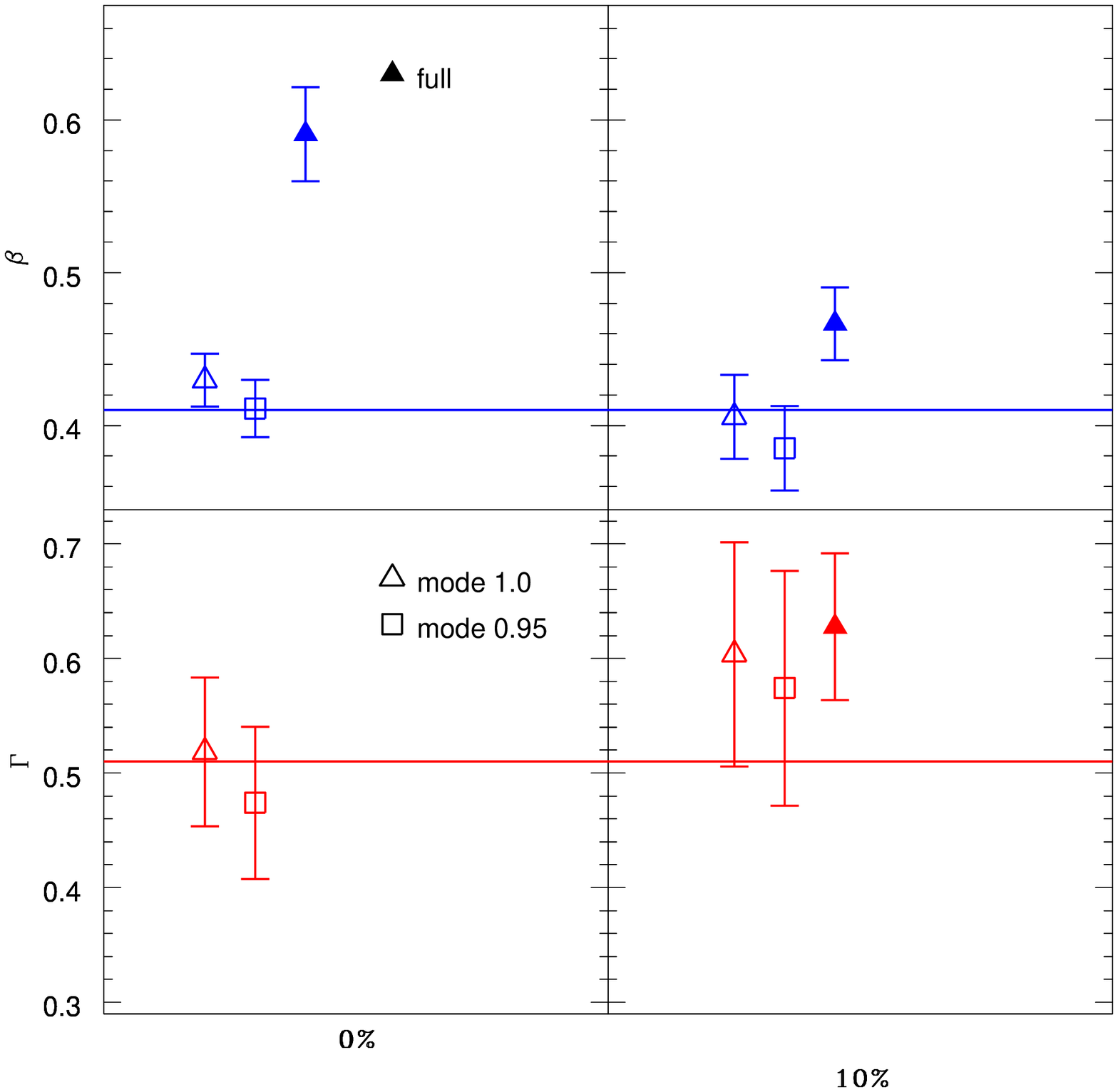}
\caption{
The same as fig~\ref{fig-2loz0} with different cosmological parameters
($\Gamma=0.51$, $\beta=0.41$).}
\label{fig-2lo-35-z0}
\end{figure}

\begin{figure}[ht]
\plotone{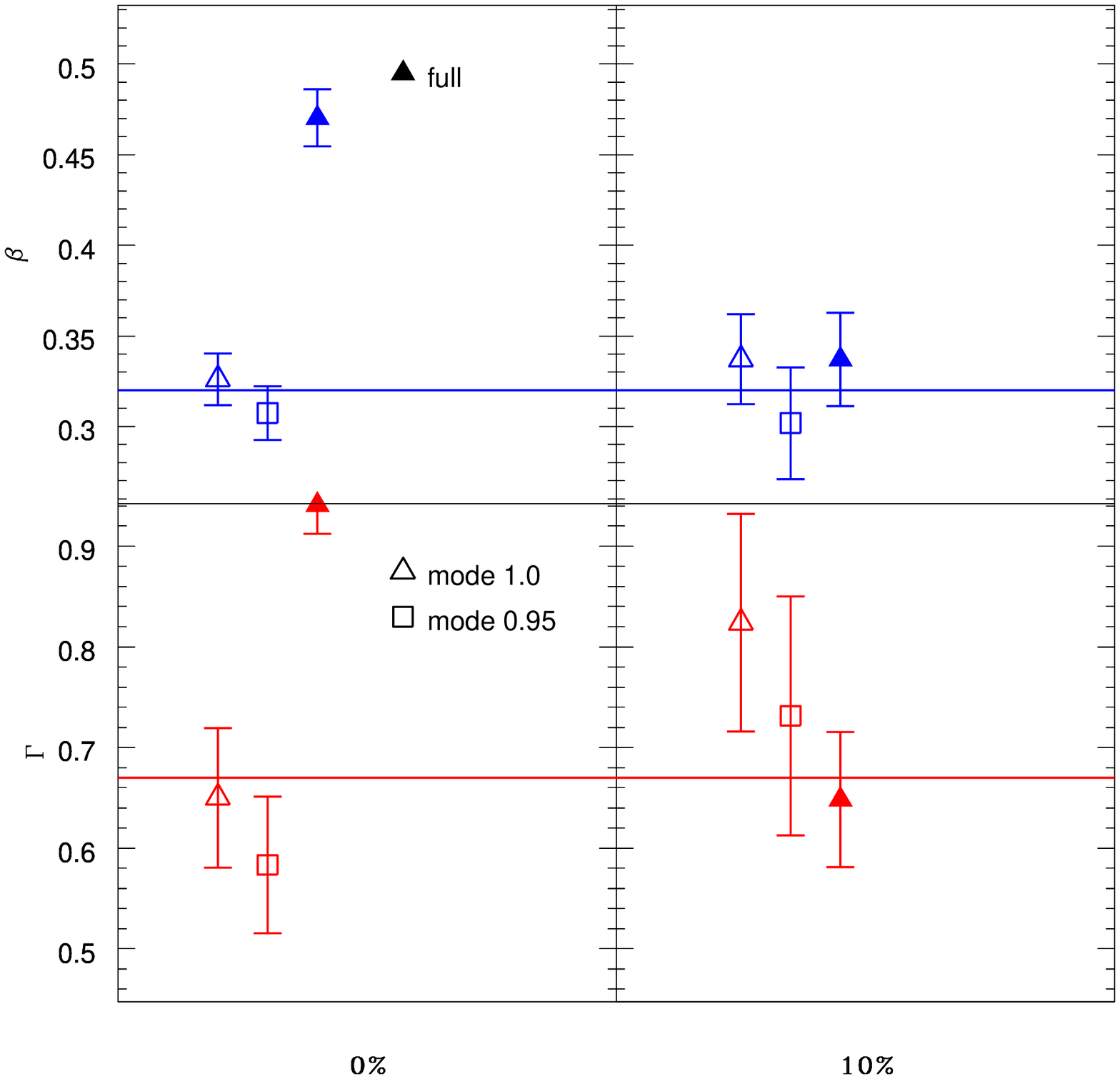}
\caption{
The same as fig~\ref{fig-2loz0} with different cosmological parameters
($\Gamma=0.667$, $\beta=0.32$).}
\label{fig-2lo-one-z0}
\end{figure}

The main purpose of the formalism we presented here and in Paper I was
to allow the removal or filtering of small scale noise while keeping the
large scale signal. To test the success of the formalism we have created
synthetic surveys from simulations with known parameters, specifically,
$\Gamma$, the CDM power spectrum shape parameter, and $\beta$, its
amplitude (the values of the parameters we simulated is given in the
table 1 below.) To compare our method with the full analysis method, we
reemphasize that the optimal moment analysis presented here allows for
two semi--independent methods of cleaning up a survey: 1) Ordering the
moments by their eigenvalues (Eq.~\ref{ordering}); and 2) Removing the
noisiest moments (Eq.~\ref{noise}). In Figs.~\ref{fig-2loz0} and
\ref{fig-2loz1}--\ref{fig-2lo-one-z0} we show the comparison between
choosing the modes least susceptible to small scale signal (open
triangles); those that are least susceptible to small scale signal {\bf
and} are not noisy (open squares); and the full analysis (that is,
keeping all moments, the usual analysis, solid triangles). We see that
the full analysis fails to recover the ``true'' parameters by a
significant amount ($\approx4\sigma$ for no errors and $>2\sigma$ for
10\% errors). In contrast, the mode analysis recovers the values of the
parameters very well, with or without the removal of the noisy
moments.

\begin{table}[ht]
\begin{center}
\begin{tabular}{||c|c|c|c|c|c|c||}\hline
$\Gamma_T$&$\beta_T$&Analysis&$<\Gamma>$&$\sigma_\Gamma$& $<\beta>$ &
$\sigma_\beta$ \\ \hline\hline 0.225 & 0.455 & mode & 0.21 $\pm$ 0.016 &
0.14 & 0.44 $\pm$ 0.009 & 0.08
\\ \cline{3-7}
      & & full & 0.41 $\pm$ 0.026 & 0.23 & 0.58 $\pm$ 0.017 & 0.15
\\ \hline
0.225 & 0.372 & mode & 0.22 $\pm$ 0.014 & 0.11 & 0.38 $\pm$ 0.008 & 0.06
\\ \cline{3-7} & & full & 0.28 $\pm$ 0.019 & 0.14 & 0.40 $\pm$ 0.010 &
0.07
\\ \hline 
0.510 & 0.410 & mode & 0.47 $\pm$ 0.066 & 0.35 & 0.41 $\pm$ 0.019 & 0.10
\\ \cline{3-7}      
      & & full & 0.90 $\pm$ 0.079 & 0.41 & 0.59 $\pm$ 0.031 & 0.16
\\ \hline
0.510 & 0.330 & mode & 0.46 $\pm$ 0.051 & 0.26 & 0.34 $\pm$ 0.015 & 0.08
\\ \cline{3-7} & & full & 0.68 $\pm$ 0.067 & 0.35 & 0.41 $\pm$ 0.019 &
0.10
\\ \hline
0.667 & 0.320 & mode & 0.58 $\pm$ 0.068 & 0.35 & 0.31 $\pm$ 0.015 & 0.08
\\ \cline{3-7} & & full & 0.94 $\pm$ 0.029 & 0.15 & 0.47 $\pm$ 0.016 &
0.08
\\ \hline 
\hline
\end{tabular}
\caption{The values of the parameters for the simulations used to
extract the catalogs we used to test our formalism. $\Gamma_T$ and
$\beta_T$ are the ``true'' values of our parameters for each
simulation. The Analysis column denotes our proposed formalism (mode)
and the full maximum likelihood analysis (full). The $<\Gamma>$ and
$<\beta>$ columns show the mean and the standard deviation of the mean
of the parameters for each simulation. The $\sigma_\Gamma$ and
$\sigma_\beta$ columns are the standard deviations for the
parameters. When comparing the maximum likelihood values, the mode
analysis does a much better job recovering to the ``true'' values then
does the full analysis.}
\end{center}
\end{table}

For each one of the models we simulated we have extracted 27 catalogs
from each simulation box, as described in Sec.~\ref{catalogs}. The
points and errorbars in the figures are the maximum likelihood mean and
standard deviation of the mean for the analysis of all catalogs. Each
one of the catalogs were analyzed using the full maximum likelihood
analysis, keeping all moments; the maximum likelihood analysis
discarding large eigenvalue modes; and the maximum likelihood mode
analysis without the noisiest moments ($\xi>0.95$).

\begin{figure}[ht]
\plotone{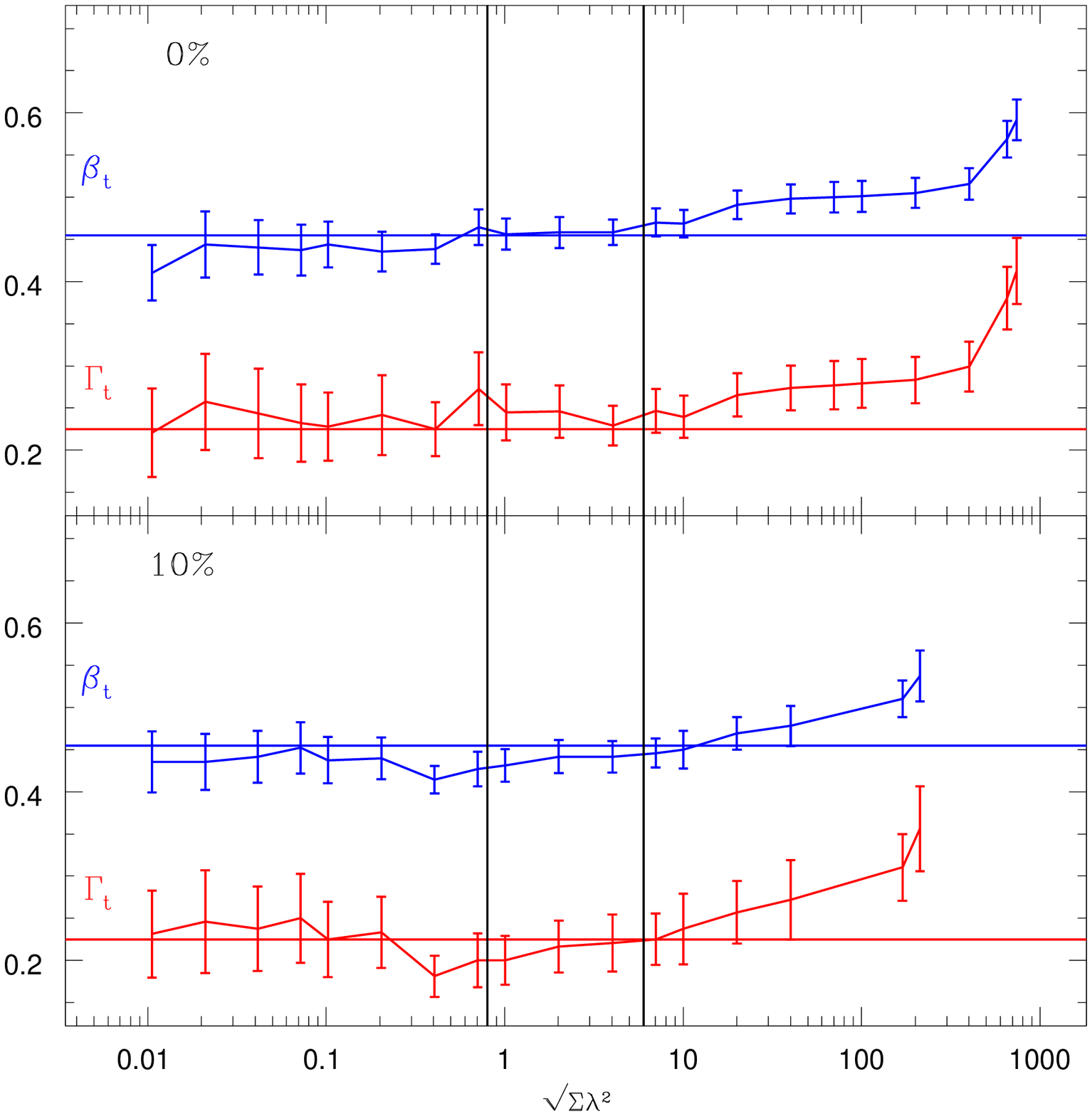}
\caption{
The mean value of the estimated parameters from 81 catalogs extracted
from the simulations (see Sec.~\ref{catalogs}) as a function of the
number of modes we keep. The top panel shows results for survey with no
errors, the bottom panel shows the results with distance errors of
10\%. It is clear that as the number of modes kept increases beyond the
criteria set, the estimators become more and more biased. The horizontal
lines are the ``true'' values of the parameters ($\Gamma=0.225$,
$\beta=0.455$). }
\label{fig-sum}
\end{figure}

In Fig.~\ref{fig-sum} we show the value of the estimated parameters as a
function of the $\Sigma\lambda^2$ (see Eq.~\ref{criteria}) where we see
that as the number of modes is increased, we get closer and closer to
the ``true'' value. When we keep more than the number of moments that
corresponds to the fulfillment of our criterion (Eq.~\ref{criteria}),
the values start diverging from the ``true'' results.  This is due to
the fact that small--scale modes that have become nonlinear are
introducing a bias.  This tendency of the full analysis to
systematically overestimate the parameter values can be seen in the
analyses done for simulations with various cosmological parameters,
Figs.~\ref{fig-2loz0} and
\ref{fig-2loz1}--\ref{fig-2lo-one-z0}.

We have experimented with the choice of $N^{\prime}$, the number of
modes to keep, as discussed in the text after Eq.~\ref{criteria}. This
choice depends on our power spectrum and more specifically on $k_{nl}$,
the wavenumber of the largest scale for which density perturbations have
become nonlinear (see Paper I).  We chose $k_{nl}$ by comparing the
linear power spectrum from the initial conditions of the simulation to
the power spectrum at the end of the run. $k_{nl}$ is where the power
spectra started to diverge. In general we found that $k_{nl}\approx0.2$,
though choosing $0.15<k_{nl}<0.4$ did not affect our results
significantly. Further, as can be seen in Fig.~\ref{fig-sum}, the
$N^{\prime}$ choice need not be finely tuned.

\begin{figure}[ht]
\plotone{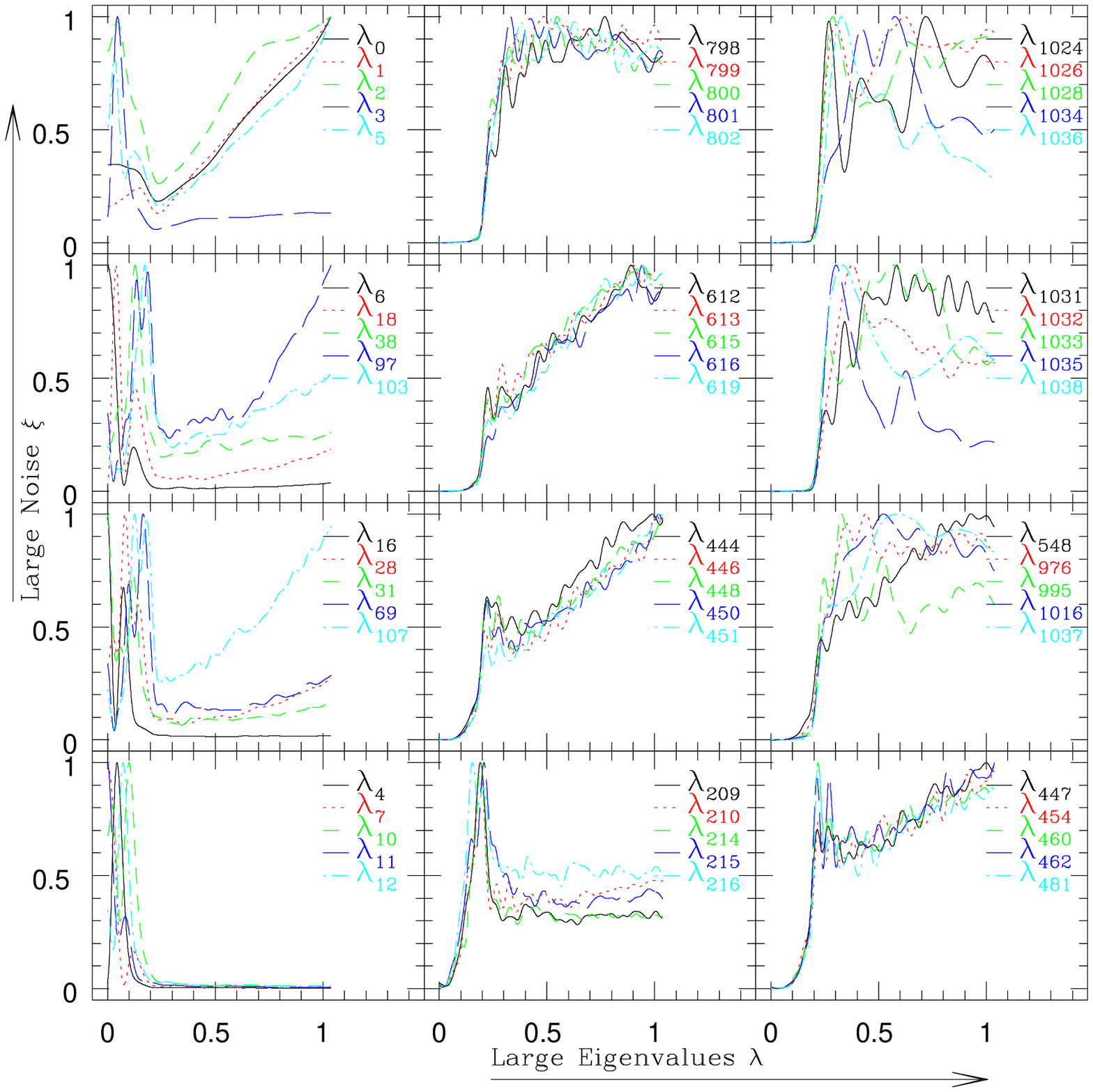}
\caption{
The window functions from top to bottom corresponding to noise in the
ranges of $0.98<\xi$, $0.95<\xi<0.98$, $0.9<\xi<0.95$ and $\xi<0.9$
respectively, and across from left low, med and high eigenvalues
$\lambda$ respectively. We can clearly see that the low eigenvalue low
noise window functions (lower left panel) probe large scale (small $k$),
whereas higher noise, larger eigenvalue window functions (up and to the
right) correspond to smaller scales probes. Further, the high noise
window functions probe scales that are hard to model as are those with
large eigenvalues.  }
\label{winfun}
\end{figure}

As was discussed in the text, the reason for the full analysis failure
to recover the ``true'' parameters when the mode analysis succeeds so
well can be shown by looking at the window functions themselves. In
fig.~\ref{winfun} we show the window functions corresponding to the five
lowest eigenvalues and lowest noise (lower left panel). Clearly, these
probe only large scales. As we move up the panels we see the window
functions with larger noise components not removed, whereas when we move
to the right we see window functions corresponding to larger
eigenvalues. Here the reasons for the particular choices for our
criteria Eqs.~(\ref{ordering}) and (\ref{criteria}) become clear. As the
eigenvalues or the noise level become large, the window functions
generally probe more small scale and less of large scale modes. Since we
are primarily interested in large scale information, discarding the
noisy, high $\lambda$ modes allows us to discard small--scale signal
that might interfere with with our analysis.

One more advantage the formalism provides is efficiency. For a catalog
of $\approx1,000$ galaxies it takes the mode analysis about one hour CPU
time, whereas it takes the full analysis about seven hours to
complete. The differences are more dramatic for larger surveys: A 5,000
galaxy catalog completes in about 30 hours with the mode analysis and
about 1,300 hours of CPU time for the full analysis. All runs were done
on the Origin2000 at the NCSA, University of Illinois,
Urbana--Champaign.

\begin{figure}[ht]
\plotone{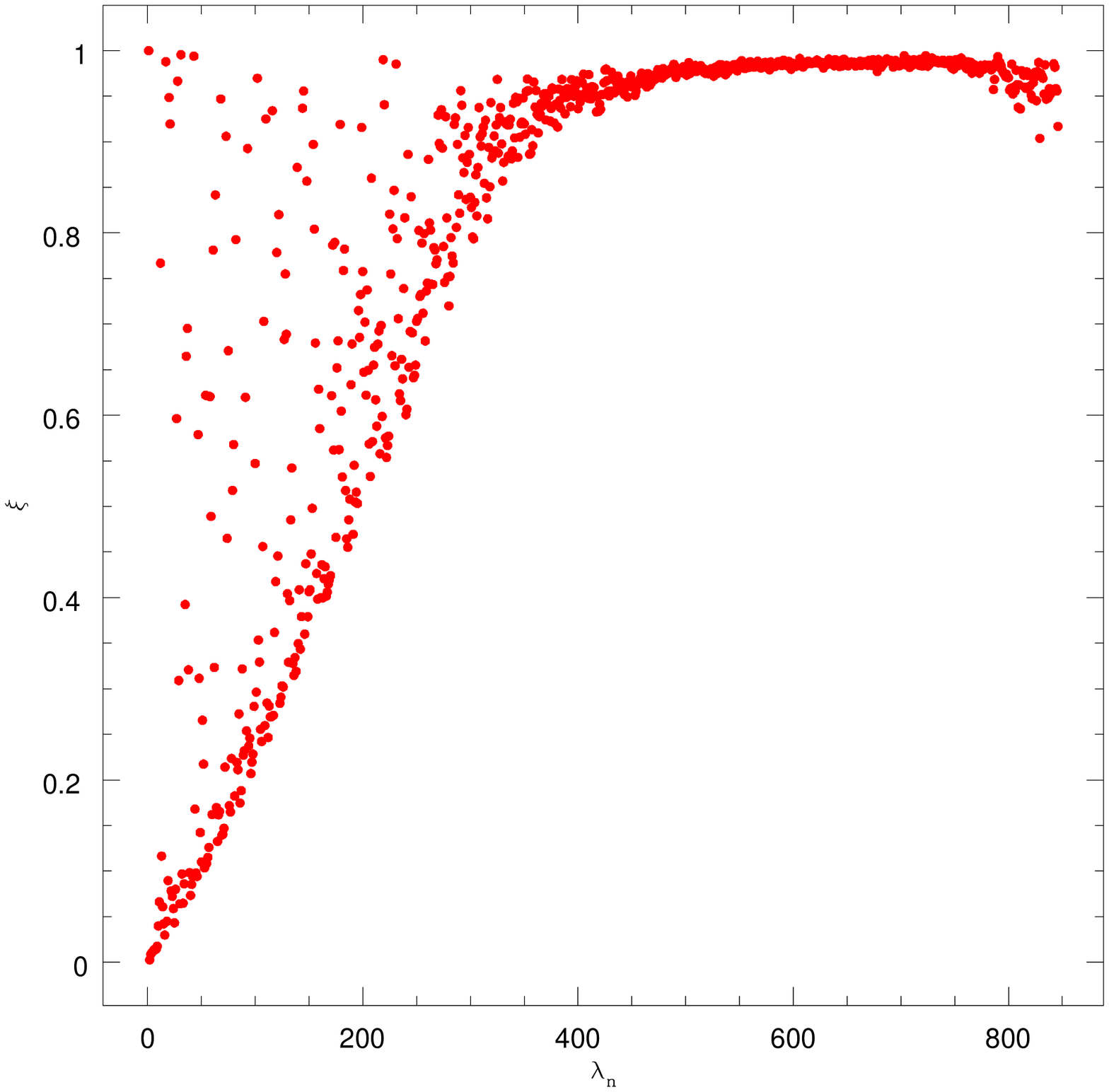}
\caption{
The noise $\xi$ (Eq.~\ref{noise}) as a function of the rank $n$ of the
eigenvalue $\lambda_n$ (Eq.~\ref{criteria}) (shown as dots). On the
average there is an excellent correlation between the rank and the noise
component.  }
\label{lambda-xi}
\end{figure}

As was mentioned in the analysis (Sec. \ref{selection}) there is a
general correlation between the $n^{\rm th}$ moment's eigenvalue
$\lambda_n$ and the noise $\xi_n$ associated with it (dots in
Fig.~\ref{lambda-xi}). As can be seen in the figure, the correlation is
not perfect, that is, there are low eigenvalues with large noise
component, but in general as we move to larger eigenvalues (large $n$)
the noise component is larger. The line in the figure represents the
running mean of the noise which shows clearly the correlations between
the noise and the eigenvalue.

\begin{figure}[ht]
\plotone{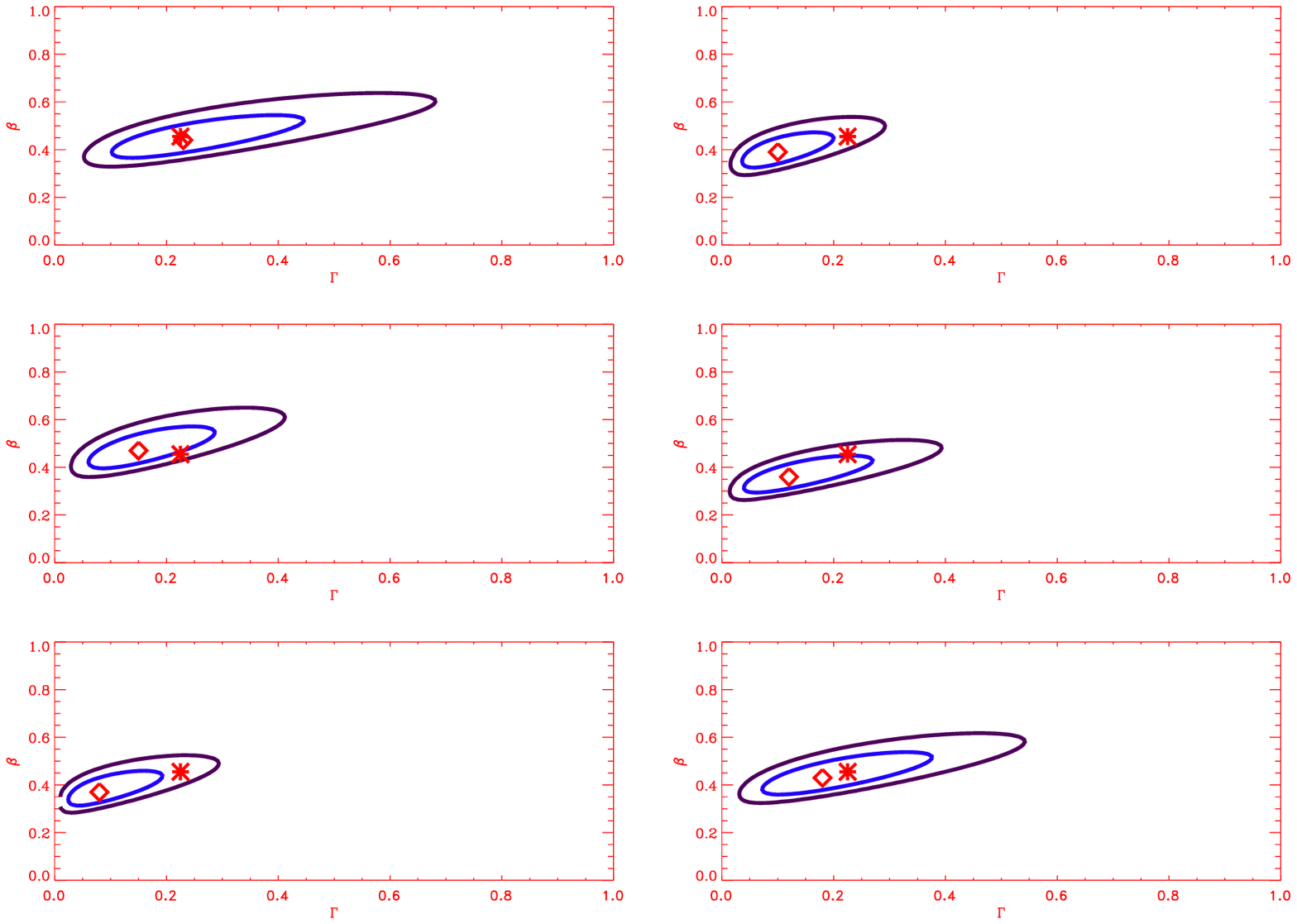}
\caption{
Maximum likelihood contours from six typical mock catalogs. The contours
are the 68\% and 94\% likelihood lines. This shows the expected
uncertainties in the analysis of one catalogs. In most cases the
uncertainties in the estimated values of the parameters $\Gamma$ and
$\beta$ are of comparable sizes to the monte carlo errorbars presented
in figures~\ref{fig-2loz0}--\ref{fig-2lo-one-z0}.  }
\label{cat-contour}
\end{figure}

In Fig.~\ref{cat-contour} we show the contours that contain 68\% and
94\% of the total likelihood for six typical catalogs. The diamond shows
the maximum likelihood results, whereas the asterix in each panel shows
the ``true'' values of the parameters. These contours allow us to
estimate the uncertainty in the maximum likelihood values obtained from
the analysis of a single catalog, as is the case when analyzing
observational data. From the figures it is clear that the uncertainties
obtained in this way are comparable to those we get from the
Monte--Carlo simulations. In general, when we try to test the
reliability of results from an observational data set, we apply our
formalism to mock catalogs extracted from N--body simulations as was
done here.  This compatibility between the uncertainties obtained in two
different ways gives us confidence that using the likelihood contours
will give us an accurate assessment of the uncertainties of our maximum
likelihood values when we apply our method to real catalogs.

\section{Conclusions}
\label{conclusions}

We have described the power and elegance of a new statistic that was
designed and formulated in order to address a crisis in the analysis of
proper distance cosmological surveys. We have shown that our formalism
mostly overcomes the problems with the traditional analysis of the
data. Whereas the full maximum likelihood analysis tends to overestimate
the values of the parameters that describe the power distribution on
large scale, our mode analysis makes very accurate estimates of these
parameters.

The formalism presented here assumes Gaussian statistics. The natural
question should be: Can the deviations from Gaussianity caused by the
collapse of perturbations interfere with the removal of small scale
power and introduce additional unpredictable biases?  As the results in
Sec.~\ref{results} indicate, deviations from Gaussianity do not have a
measurable effect and the effectiveness of filtering small--scale power
is unbiased. Further, we have explored in detail such issues as moment
and noise selection, the window functions' effectiveness and criteria
for which modes to keep.

As was shown in Paper I and in more detail here, the formalism we
presented is highly adaptive and versatile. It can be applied surveys
with any geometry and density, and since it retains maximum information
should be particularly useful for sparse data such as that obtained in
cluster peculiar velocity surveys.  Overall, we consider this method to
be a significant improvement over previous methods used for the analysis
of peculiar velocity data.

\acknowledgments

We wish to thank Jim Fry, Roman Juszkiewicz and Avishai Dekel for
illuminating conversations.  HAF and ALM wish to acknowledge support
from the National Science Foundation under grant number AST--0070702,
the University of Kansas General Research Fund and the National Center
for Supercomputing Applications for allocation of computer time. This
research has been partially supported by the Lady Davis Foundation at
the Hebrew University, Jerusalem, Israel and by the Institute of
Theoretical Physics at the Technion, Haifa, Israel.

\end{document}